 %
 %
%
%
%
\def\unredoffs{} \def\redoffs{\voffset=-.16truein\hoffset=-.49truein}
\def\speclscape{}
%
%
%
%
\newbox\leftpage \newdimen\fullhsize \newdimen\hstitle \newdimen\hsbody
\tolerance=1000\hfuzz=2pt
\catcode`\@=11 
\def\bigans{b }
\let\answ\bigans
\ifx\answ\bigans\message{(This will come out unreduced.}
\magnification=1200\unredoffs\baselineskip=16pt plus 2pt minus 1pt
\hsbody=\hsize \hstitle=\hsize 
\else\message{(This will be reduced.} \let\l@r=L
\magnification=1000\baselineskip=16pt plus 2pt minus 1pt \vsize=7truein
\redoffs \hstitle=8truein\hsbody=4.75truein\fullhsize=10truein\hsize=\hsbody
\output={\ifnum\pageno=0 
  \shipout\vbox{\speclscape{\hsize\fullhsize\makeheadline}
    \hbox to \fullhsize{\hfill\pagebody\hfill}}\advancepageno
  \else
  \almostshipout{\leftline{\vbox{\pagebody\makefootline}}}\advancepageno
  \fi}
\def\almostshipout#1{\if L\l@r \count1=1 \message{[\the\count0.\the\count1]}
      \global\setbox\leftpage=#1 \global\let\l@r=R
 \else \count1=2
  \shipout\vbox{\speclscape{\hsize\fullhsize\makeheadline}
      \hbox to\fullhsize{\box\leftpage\hfil#1}}  \global\let\l@r=L\fi}
\fi
%
\newcount\yearltd\yearltd=\year\advance\yearltd by -1900

\def\Date#1{\vfill\leftline{#1}\tenpoint\supereject\global\hsize=\hsbody%
\footline={\hss\tenrm\folio\hss}}
%

\def\draftmode{\message{ DRAFTMODE }\def\draftdate{{\rm preliminary draft:
\number\month/\number\day/\number\yearltd\ \ \hourmin}}%
\headline={\hfil\draftdate}\writelabels\baselineskip=20pt plus 2pt minus 2pt
 {\count255=\time\divide\count255 by 60 \xdef\hourmin{\number\count255}
  \multiply\count255 by-60\advance\count255 by\time
  \xdef\hourmin{\hourmin:\ifnum\count255<10 0\fi\the\count255}}}
\def\nolabels{\def\wrlabeL##1{}\def\eqlabeL##1{}\def\reflabeL##1{}}
\def\writelabels{\def\wrlabeL##1{\leavevmode\vadjust{\rlap{\smash%
{\line{{\escapechar=` \hfill\rlap{\sevenrm\hskip.03in\string##1}}}}}}}%
\def\eqlabeL##1{{\escapechar-1\rlap{\sevenrm\hskip.05in\string##1}}}%
\def\reflabeL##1{\noexpand\llap{\noexpand\sevenrm\string\string\string##1}}}
\nolabels
%
\global\newcount\secno \global\secno=0
\global\newcount\meqno \global\meqno=1
\def\newsec#1{\global\advance\secno by1\message{(\the\secno. #1)}
\global\subsecno=0\eqnres@t\noindent{\bf\the\secno. #1}
\writetoca{{\secsym} {#1}}\par\nobreak\medskip\nobreak}
\def\eqnres@t{\xdef\secsym{\the\secno.}\global\meqno=1\bigbreak\bigskip}
\def\sequentialequations{\def\eqnres@t{\bigbreak}}\xdef\secsym{}
\global\newcount\subsecno \global\subsecno=0
\def\subsec#1{\global\advance\subsecno by1\message{(\secsym\the\subsecno. #1)}
\ifnum\lastpenalty>9000\else\bigbreak\fi
\noindent{\it\secsym\the\subsecno. #1}\writetoca{\string\quad
{\secsym\the\subsecno.} {#1}}\par\nobreak\medskip\nobreak}
\def\appendix#1#2{\global\meqno=1\global\subsecno=0\xdef\secsym{\hbox{#1.}}
\bigbreak\bigskip\noindent{\bf Appendix #1. #2}\message{(#1. #2)}
\writetoca{Appendix {#1.} {#2}}\par\nobreak\medskip\nobreak}
%
%
\def\eqnn#1{\xdef #1{(\secsym\the\meqno)}\writedef{#1\leftbracket#1}%
\global\advance\meqno by1\wrlabeL#1}
\def\eqna#1{\xdef #1##1{\hbox{$(\secsym\the\meqno##1)$}}
\writedef{#1\numbersign1\leftbracket#1{\numbersign1}}%
\global\advance\meqno by1\wrlabeL{#1$\{\}$}}
\def\eqn#1#2{\xdef #1{(\secsym\the\meqno)}\writedef{#1\leftbracket#1}%
\global\advance\meqno by1$$#2\eqno#1\eqlabeL#1$$}
%
\newskip\footskip\footskip14pt plus 1pt minus 1pt 
\def\footnotefont{\tenpoint}\def\f@t#1{\footnotefont #1\@foot}
\def\f@@t{\baselineskip\footskip\bgroup\footnotefont\aftergroup\@foot\let\next}
\setbox\strutbox=\hbox{\vrule height9.5pt depth4.5pt width0pt}
\global\newcount\ftno \global\ftno=0
\def\foot{\global\advance\ftno by1\footnote{$^{\the\ftno}$}}
%
\newwrite\ftfile
\def\footend{\def\foot{\global\advance\ftno by1\chardef\wfile=\ftfile
$^{\the\ftno}$\ifnum\ftno=1\immediate\openout\ftfile=foots.tmp\fi%
\immediate\write\ftfile{\noexpand\smallskip%
\noexpand\item{f\the\ftno:\ }\pctsign}\findarg}%
\def\footatend{\vfill\eject\immediate\closeout\ftfile{\parindent=20pt
\centerline{\bf Footnotes}\nobreak\bigskip\input foots.tmp }}}
\def\footatend{}
%
%
\global\newcount\refno \global\refno=1
\newwrite\rfile
\def\ref{[\the\refno]\nref}
\def\nref#1{\xdef#1{[\the\refno]}\writedef{#1\leftbracket#1}%
\ifnum\refno=1\immediate\openout\rfile=refs.tmp\fi
\global\advance\refno by1\chardef\wfile=\rfile\immediate
\write\rfile{\noexpand\item{#1\ }\reflabeL{#1\hskip.31in}\pctsign}\findarg}
\def\findarg#1#{\begingroup\obeylines\newlinechar=`\^^M\pass@rg}
{\obeylines\gdef\pass@rg#1{\writ@line\relax #1^^M\hbox{}^^M}%
\gdef\writ@line#1^^M{\expandafter\toks0\expandafter{\striprel@x #1}%
\edef\next{\the\toks0}\ifx\next\em@rk\let\next=\endgroup\else\ifx\next\empty%
\else\immediate\write\wfile{\the\toks0}\fi\let\next=\writ@line\fi\next\relax}}
\def\striprel@x#1{} \def\em@rk{\hbox{}}
\def\lref{\begingroup\obeylines\lr@f}
\def\lr@f#1#2{\gdef#1{\ref#1{#2}}\endgroup\unskip}

\def\addref#1{\immediate\write\rfile{\noexpand\item{}#1}} 
\def\listrefs{\footatend\vfill\supereject\immediate\closeout\rfile\writestoppt
\baselineskip=14pt\centerline{{\bf References}}\bigskip{\frenchspacing%
\parindent=20pt\escapechar=` \input refs.tmp\vfill\eject}\nonfrenchspacing}
\def\startrefs#1{\immediate\openout\rfile=refs.tmp\refno=#1}
\def\xref{\expandafter\xr@f}\def\xr@f[#1]{#1}
\def\refs#1{\count255=1[\r@fs #1{\hbox{}}]}
\def\r@fs#1{\ifx\und@fined#1\message{reflabel \string#1 is undefined.}%
\nref#1{need to supply reference \string#1.}\fi%
\vphantom{\hphantom{#1}}\edef\next{#1}\ifx\next\em@rk\def\next{}%
\else\ifx\next#1\ifodd\count255\relax\xref#1\count255=0\fi%
\else#1\count255=1\fi\let\next=\r@fs\fi\next}
%

%
\newwrite\ffile\global\newcount\figno \global\figno=1
\def\fig{fig.~\the\figno\nfig}
\def\nfig#1{\xdef#1{fig.~\the\figno}%
\writedef{#1\leftbracket fig.\noexpand~\the\figno}%
\ifnum\figno=1\immediate\openout\ffile=figs.tmp\fi\chardef\wfile=\ffile%
\immediate\write\ffile{\noexpand\medskip\noexpand\item{Fig.\ \the\figno. }
\reflabeL{#1\hskip.55in}\pctsign}\global\advance\figno by1\findarg}
\def\xfig{\expandafter\xf@g}\def\xf@g fig.\penalty\@M\ {}
\def\figs#1{figs.~\f@gs #1{\hbox{}}}
\def\f@gs#1{\edef\next{#1}\ifx\next\em@rk\def\next{}\else
\ifx\next#1\xfig #1\else#1\fi\let\next=\f@gs\fi\next}
\newwrite\lfile
{\escapechar-1\xdef\pctsign{\string\%}\xdef\leftbracket{\string\{}
\xdef\rightbracket{\string\}}\xdef\numbersign{\string\#}}

\def\writestop{\def\writestoppt{\immediate\write\lfile{\string\pageno%
\the\pageno\string\startrefs\leftbracket\the\refno\rightbracket%
\string\def\string\secsym\leftbracket\secsym\rightbracket%
\string\secno\the\secno\string\meqno\the\meqno}\immediate\closeout\lfile}}
\def\writestoppt{}\def\writedef#1{}
\def\seclab#1{\xdef #1{\the\secno}\writedef{#1\leftbracket#1}\wrlabeL{#1=#1}}
\def\subseclab#1{\xdef #1{\secsym\the\subsecno}%
\writedef{#1\leftbracket#1}\wrlabeL{#1=#1}}
\newwrite\tfile \def\writetoca#1{}
\def\leaderfill{\leaders\hbox to 1em{\hss.\hss}\hfill}
\def\writetoc{\immediate\openout\tfile=toc.tmp
   \def\writetoca##1{{\edef\next{\write\tfile{\noindent ##1
   \string\leaderfill {\noexpand\number\pageno} \par}}\next}}}

%
\catcode`\@=12 
%
\edef\tfontsize{\ifx\answ\bigans scaled\magstep3\else scaled\magstep4\fi}
\font\titlerm=cmr10 \tfontsize \font\titlerms=cmr7 \tfontsize
 \tfontsize \font\titlei=cmmi10 \tfontsize
\font\titleis=cmmi7 \tfontsize \font\titleiss=cmmi5 \tfontsize
\font\titlesy=cmsy10 \tfontsize \font\titlesys=cmsy7 \tfontsize
\font\titlesyss=cmsy5 \tfontsize  \tfontsize
\skewchar\titlei='177 \skewchar\titleis='177 \skewchar\titleiss='177
\skewchar\titlesy='60 \skewchar\titlesys='60 \skewchar\titlesyss='60
 \ifx\answ\bigans\else scaled\magstep1\fi
\ifx\answ\bigans\else

 \font\absi=cmmi10 scaled\magstep1
\font\absis=cmmi7 scaled\magstep1 \font\absiss=cmmi5 scaled\magstep1
\font\abssy=cmsy10 scaled\magstep1 \font\abssys=cmsy7 scaled\magstep1
\font\abssyss=cmsy5 scaled\magstep1 
\skewchar\absi='177 \skewchar\absis='177 \skewchar\absiss='177
\skewchar\abssy='60 \skewchar\abssys='60 \skewchar\abssyss='60
\fi
\def\tenpoint{\def\rm{\fam0\tenrm}
\textfont0=\tenrm \scriptfont0=\sevenrm \scriptscriptfont0=\fiverm
\textfont1=\teni  \scriptfont1=\seveni  \scriptscriptfont1=\fivei
\textfont2=\tensy \scriptfont2=\sevensy \scriptscriptfont2=\fivesy
\textfont\itfam=\tenit \def\it{\fam\itfam\tenit}\def\footnotefont{\ninepoint}%
\textfont\bffam=\tenbf \def\bf{\fam\bffam\tenbf}\def\sl{\fam\slfam\tensl}\rm}
\font\ninerm=cmr9 \font\sixrm=cmr6 \font\ninei=cmmi9 \font\sixi=cmmi6
\font\ninesy=cmsy9 \font\sixsy=cmsy6 \font\ninebf=cmbx9
\font\nineit=cmti9 \font\ninesl=cmsl9 \skewchar\ninei='177
\skewchar\sixi='177 \skewchar\ninesy='60 \skewchar\sixsy='60
\def\ninepoint{\def\rm{\fam0\ninerm}
\textfont0=\ninerm \scriptfont0=\sixrm \scriptscriptfont0=\fiverm
\textfont1=\ninei \scriptfont1=\sixi \scriptscriptfont1=\fivei
\textfont2=\ninesy \scriptfont2=\sixsy \scriptscriptfont2=\fivesy
\textfont\itfam=\ninei \def\it{\fam\itfam\nineit}\def\sl{\fam\slfam\ninesl}%
\textfont\bffam=\ninebf \def\bf{\fam\bffam\ninebf}\rm}
%
%
\def\noblackbox{\overfullrule=0pt}
\hyphenation{anom-aly anom-alies coun-ter-term coun-ter-terms}
\def\inv{^{\raise.15ex\hbox{${\scriptscriptstyle -}$}\kern-.05em 1}}

\def\Dsl{\,\raise.15ex\hbox{/}\mkern-13.5mu D} 
\def\dsl{\raise.15ex\hbox{/}\kern-.57em\partial}

\def\lspace{\ifx\answ\bigans{}\else\qquad\fi}
\def\lbspace{\ifx\answ\bigans{}\else\hskip-.2in\fi} 
\def\boxeqn#1{\vcenter{\vbox{\hrule\hbox{\vrule\kern3pt\vbox{\kern3pt
        \hbox{${\displaystyle #1}$}\kern3pt}\kern3pt\vrule}\hrule}}}
\def\mbox#1#2{\vcenter{\hrule \hbox{\vrule height#2in
                \kern#1in \vrule} \hrule}}  
%

\def\e#1{{\rm e}^{^{\textstyle#1}}}

\def\darr#1{\raise1.5ex\hbox{$\leftrightarrow$}\mkern-16.5mu #1}

\def\roughly#1{\raise.3ex\hbox{$#1$\kern-.75em\lower1ex\hbox{$\sim$}}}

 %
\catcode`@=11
\def\rlx{\relax\leavevmode}                  
 %
 %
 %
\font\tenmib=cmmib10
\font\sevenmib=cmmib10 at 7pt 
\font\fivemib=cmmib10 at 5pt  
\font\tenbsy=cmbsy10
\font\sevenbsy=cmbsy10 at 7pt 
\font\fivebsy=cmbsy10 at 5pt  
\def\BMfont{\textfont0\tenbf \scriptfont0\sevenbf
                              \scriptscriptfont0\fivebf
            \textfont1\tenmib \scriptfont1\sevenmib
                               \scriptscriptfont1\fivemib
            \textfont2\tenbsy \scriptfont2\sevenbsy
                               \scriptscriptfont2\fivebsy}
\def\BM#1{\rlx\ifmmode\mathchoice
                      {\hbox{$\BMfont#1$}}
                      {\hbox{$\BMfont#1$}}
                      {\hbox{$\scriptstyle\BMfont#1$}}
                      {\hbox{$\scriptscriptstyle\BMfont#1$}}
                 \else{$\BMfont#1$}\fi}
 %
 %
 %
 %
\def\inbar{\vrule height1.5ex width.4pt depth0pt}
\def\sinbar{\vrule height1ex width.35pt depth0pt}
\def\ssinbar{\vrule height.7ex width.3pt depth0pt}
\font\cmss=cmss10
\font\cmsss=cmss10 at 7pt
\def\ZZ{\rlx\leavevmode
             \ifmmode\mathchoice
                    {\hbox{\cmss Z\kern-.4em Z}}
                    {\hbox{\cmss Z\kern-.4em Z}}
                    {\lower.9pt\hbox{\cmsss Z\kern-.36em Z}}
                    {\lower1.2pt\hbox{\cmsss Z\kern-.36em Z}}
               \else{\cmss Z\kern-.4em Z}\fi}
\def\Ik{\rlx{\rm I\kern-.18em k}}  
\def\IC{\rlx\leavevmode
             \ifmmode\mathchoice
                    {\hbox{\kern.33em\inbar\kern-.3em{\rm C}}}
                    {\hbox{\kern.33em\inbar\kern-.3em{\rm C}}}
                    {\hbox{\kern.28em\sinbar\kern-.25em{\sevenrm C}}}
                    {\hbox{\kern.25em\ssinbar\kern-.22em{\fiverm C}}}
             \else{\hbox{\kern.3em\inbar\kern-.3em{\rm C}}}\fi}
\def\IP{\rlx{\rm I\kern-.18em P}}
\def\IR{\rlx{\rm I\kern-.18em R}}
\def\Ione{\rlx{\rm 1\kern-2.7pt l}}
 %
 %

 %

\def\intem#1{\par\leavevmode%
              \llap{\hbox to\parindent{\hss{#1}\hfill~}}\ignorespaces}
 %


 %
\newskip\humongous \humongous=0pt plus 1000pt minus 1000pt   
\def\caja{\mathsurround=0pt}
\newif\ifdtup
 %
\def\eqalign#1{\,\vcenter{\openup2\jot \caja
     \ialign{\strut \hfil$\displaystyle{##}$&$
      \displaystyle{{}##}$\hfil\crcr#1\crcr}}\,}
 %

 %
\def\panorama{\global\dtuptrue \openup2\jot \caja
     \everycr{\noalign{\ifdtup \global\dtupfalse
      \vskip-\lineskiplimit \vskip\normallineskiplimit
      \else \penalty\interdisplaylinepenalty \fi}}}
 %
\def\eqalignno#1{\panorama \tabskip=\humongous
     \halign to\displaywidth{\hfil$\displaystyle{##}$
      \tabskip=0pt&$\displaystyle{{}##}$\hfil
       \tabskip=\humongous&\llap{$##$}\tabskip=0pt\crcr#1\crcr}}
 %

 %

 %

 %
 %
 %
 %
   \let\SS=\S       
\def\,{\hskip1.5pt}           
 %

\let\c=\chi
                    
\let\e=\epsilon     
\let\f=\phi         \let\vf=\varphi

\let\j=\psi                                      \let\J=\Psi

                                   \let\L=\Lambda

\let\p=\pi                         
\let\q=\theta                   
         
                   \let\S=\Sigma

 %
 %
\def\Box{\sqcap\llap{$\sqcup$}}
\def\lapp{\lower.4ex\hbox{\rlap{$\sim$}} \raise.4ex\hbox{$<$}}
\def\gapp{\lower.4ex\hbox{\rlap{$\sim$}} \raise.4ex\hbox{$>$}}
\def\con{\ifmmode\raise.1ex\hbox{\bf*}
          \else\raise.1ex\hbox{\bf*}\fi}
\def\bo{{\raise.15ex\hbox{\large$\Box\kern-.39em$}}}

\def\dual{\relax\leavevmode\lower.9ex\hbox{\titlerms*}}
\def\define{\buildrel\rm def\over =}

\let\8=\otimes
 %
 %
 %
 %

\let\2=\underline

 %
\def\dt#1{{\buildrel{\smash{\lower1pt\hbox{.}}}\over{#1}}}

\font\eightrm=cmr8
\def\6(#1){\relax\leavevmode\hbox{\eightrm(}#1\hbox{\eightrm)}}
\def\0#1{\relax\ifmmode\mathaccent"7017{#1}     
                \else\accent23#1\relax\fi}      
\def\7#1#2{{\mathop{\null#2}\limits^{#1}}}      
\def\5#1#2{{\mathop{\null#2}\limits_{#1}}}      
 %

\def\ket#1{\left| #1\right\rangle}
\def\V#1{\langle#1\rangle}

 %

 %

 %

 %
\newbox\t@b@x
\def\rightarrowfill{$\m@th \mathord- \mkern-6mu
     \cleaders\hbox{$\mkern-2mu \mathord- \mkern-2mu$}\hfill
      \mkern-6mu \mathord\rightarrow$}
\def\tooo#1{\setbox\t@b@x=\hbox{$\scriptstyle#1$}%
             \mathrel{\mathop{\hbox to\wd\t@b@x{\rightarrowfill}}%
              \limits^{#1}}\,}
\def\leftarrowfill{$\m@th \mathord\leftarrow \mkern-6mu
     \cleaders\hbox{$\mkern-2mu \mathord- \mkern-2mu$}\hfill
      \mkern-6mu \mathord-$}
\def\froo#1{\setbox\t@b@x=\hbox{$\scriptstyle#1$}%
             \mathrel{\mathop{\hbox to\wd\t@b@x{\leftarrowfill}}%
              \limits^{#1}}\,}
 %
\def\frac#1#2{{#1\over#2}}
\def\frc#1#2{\relax\ifmmode{\textstyle{#1\over#2}} 
                    \else$#1\over#2$\fi}           
\def\inv#1{\frc{1}{#1}}                            
 %
\def\Claim#1#2#3{\bigskip\begingroup%
                  \xdef #1{\secsym\the\meqno}%
                   \writedef{#1\leftbracket#1}%
                    \global\advance\meqno by1\wrlabeL#1%
                     \noindent{\bf#2}\,#1{}\,:~\sl#3\vskip1mm\endgroup}

\def\QED{\rlx\hfill$\Box$\kern-7pt\raise3pt\hbox{$\surd$}\bigskip}
 %
 %

 %
\def\muthstrut{\vphantom1}
\def\mutrix#1{\null\,\vcenter{\normalbaselines\m@th
        \ialign{\hfil$##$\hfil&&~\hfil$##$\hfill\crcr
            \muthstrut\crcr\noalign{\kern-\baselineskip}
            #1\crcr\muthstrut\crcr\noalign{\kern-\baselineskip}}}\,}

 %
\def\YT#1#2{\vcenter{\hbox{\vbox{\baselineskip0pt\parskip=\medskipamount%
             \def\Box{$\sqcap\llap{$\sqcup$}$\kern-1.2pt}%
              \def\Z{\hfil\vskip-5.8pt}\lineskiplimit0pt\lineskip0pt%
               \setbox0=\hbox{#1}\hsize\wd0\parindent=0pt#2}\,}}}
\def\EU{\rlx\ifmmode \c_{{}_E} \else$\c_{{}_E}$\fi}
\def\TM{\rlx\ifmmode {\cal T_M} \else$\cal T_M$\fi}
\def\TW{\rlx\ifmmode {\cal T_W} \else$\cal T_W$\fi}
\def\CM{\rlx\ifmmode {\cal T\rlap{\bf*}\!\!_M}
             \else$\cal T\rlap{\bf*}\!\!_M$\fi}
\def\hm#1#2{\rlx\ifmmode H^{#1}({\cal M},{#2})
                 \else$H^{#1}({\cal M},{#2})$\fi}
\def\CP#1{\rlx\ifmmode\IP^{#1}\else\IP$^{#1}$\fi}
\def\cP#1{\rlx\ifmmode\IC{\rm P}^{#1}\else$\IC{\rm P}^{#1}$\fi}

\def\sll#1{\rlx\rlap{\,\raise1pt\hbox{/}}{#1}}
\def\Sll#1{\rlx\rlap{\,\kern.6pt\raise1pt\hbox{/}}{#1}\kern-.6pt}
%

 %
 %
\def\ie{\hbox{\it i.e.}}        

\def\CY{Calabi-\kern-.2em Yau}

\def\3{\ifmmode\ldots\else$\ldots$\fi}
\def\Z{\hfil\break\rlx\hbox{}\quad}
\def\3{\ifmmode\ldots\else$\ldots$\fi}
\def\?{d\kern-.3em\raise.64ex\hbox{-}}           
\def\9{\raise.43ex\hbox{-}\kern-.37em D}         
\def\ping{\nobreak\par\centerline{---$\circ$---}\goodbreak\bigskip}
 %
 %

 %

 %

 %
 %
 %
\baselineskip=13.0861pt plus2pt minus1pt
\parskip=\medskipamount
\let\ft=\foot
\noblackbox
\def\SaveTimber{\abovedisplayskip=1.5ex plus.3ex minus.5ex
                \belowdisplayskip=1.5ex plus.3ex minus.5ex
                \abovedisplayshortskip=.2ex plus.2ex minus.4ex
                \belowdisplayshortskip=1.5ex plus.2ex minus.4ex
                \baselineskip=12pt plus1pt minus.5pt
 \parskip=\smallskipamount
 \def\ft##1{\unskip\,\begingroup\footskip9pt plus1pt minus1pt\setbox%
             \strutbox=\hbox{\vrule height6pt depth4.5pt width0pt}%
              \global\advance\ftno by1\footnote{$^{\the\ftno)}$}{##1}%
               \endgroup}
 \def\listrefs{\footatend\vfill\immediate\closeout\rfile%
                \writestoppt\baselineskip=10pt%
                 \centerline{{\bf References}}%
                  \bigskip{\frenchspacing\parindent=20pt\escapechar=` %
                   \rightskip=0pt plus4em\spaceskip=.3333em%
                    \input refs.tmp\vfill\eject}\nonfrenchspacing}}
 %
\def\Afour{\ifx\answ\bigans
            \hsize=16.5truecm\vsize=24.7truecm
             \else
              \hsize=24.7truecm\vsize=16.5truecm
               \fi}
\catcode`@=12
 %
 %
\def\rd{{\rm d}}
\def\\{\hfill\break}
\def\jt#1{\item{{\bf#1}}}
 %
 %
\pageno=0
\footline{\ifnum\pageno=0{}\else{\hss\tenrm--\,\folio\,--\hss}\fi}
\noblackbox
\vglue0pt\vfill
\vglue15mm
 \centerline{\titlerm   Quantum Mechanics is Either}          \vskip4mm
 \centerline{\titlerm   Non-Linear Or Non-Introspective}      \vskip10mm
 \centerline{\titlerms  Tristan H\"ubsch\footnote{$^{\spadesuit}$}
       {On leave from the Institut Rudjer Bo\v{s}kovi\'c, Zagreb,
        Croatia.\\ Supported by the US Department of Energy grant
        DE-FG02-94ER-40854.}}                                 \vskip-1mm
 \centerline{Department of Physics and Astronomy}             \vskip-1mm
 \centerline{Howard University, Washington, DC 20059}         \vskip-1mm
 \centerline{\tt thubsch\,@\,howard.edu}                      \vglue+4mm
 \rightline{{\ninepoint All paradoxes can be paradoctored.
                                                      \sl--R.A.~Heinlein}}
\vfill

\centerline{ABSTRACT}\vskip2mm
\vbox{\rightskip=4.5em\leftskip=\rightskip\baselineskip=12pt\noindent
 The measurement conundrum seems to have plagued quantum mechanics for so
long that impressions of an inconsistency amongst its axioms have spawned.
 A demonstration that such purported inconsistency is fictitious may then
be in order and is presented here.
 An exclusion principle of sorts emerges, stating that quantum mechanics
cannot be simultaneously linear and introspective (self-observing).}

\Date{December 1997\hfill}
\vfill\eject
\footline{\hss\tenrm--\,\folio\,--\hss}
 %
 %
\lref\rAlbert{D.Z.~Albert: {\sl Quantum Mechanics and Experience},\Z
       (Harvard University Press, Cambridge, MA, 1992).}

\lref\rAuyang{S.Y~Auyang: {\sl How is Quantum Field Theory Possible?},\Z
       (Oxford University Press, Oxford, 1995).}

\lref\rCTDL{C.~Cohen-Tannoudji, B.~Diu and F.~Lalo\"e: {\sl Quantum
       Mechanics},\Z (John Wiley and Sons, New York, 1977).}

\lref\rPark{D.~Park: {\sl Introduction to the Quantum Theory} 3rd ed.,\Z
       (McGraw-Hill, New York, 1992).}

 %
 %
\SaveTimber
\newsec{Axioms and Assumptions}\noindent
Although close to becoming a centenarian, quantum mechanics still has
adolescent (although not obviously just cosmetic) problems, most notably
exemplified by the conundrum known as the ``quantum measurement
problem''. The conundrum has been considered from very diverse points of
view and phrased in many different ways, including the claim of
contradiction between two of its axioms~\rAlbert, hence an inherent
inconsistency of quantum mechanics as a scientific theory.

The purpose of this articlet is to show that this particular (apparent)
contradiction stems from a slight and subtle but serious misinterpretation
of the axioms --- a misinterpretation which however appears to be too well
hidden and all too frequent to be easily dismissed as trivial. On exposing
this misinterpretation, an avenue seems to open for a possible and perhaps
interesting resolution of the ``quantum measurement problem''. Details of
this quest are however beyond our present scope.

The routine maneuver in some relevant applications is then seen to confirm
the main result, stated in the title. While this will surprise no seasoned
practitioner, a clear and explicit statement is to the best of knowledge
of the present author nowhere to be found in print, and may therefore turn
out to be welcome.
\ping

Over the years, {\it one} collection of axioms\ft{The `axioms' and
`theorems' of any system may always be reorganized so as to swap
a `theorem' with an `axiom'---provided the rules of deduction allow the
demoted `axiom' to be derived from the new circle of `axioms'.} has become
more frequently quoted than any other. For the sake of completeness, they
are~\rCTDL\ (with slight adaptation):
\jt{1.} At any given time, $t$, the state of a physical system is
defined by specifying a state-function (ket), $\ket{\j}$, belonging
to the state set $\cal E$.
\jt{2.} Every quantity $\cal A$ which can be measured (at least in
principle) is ascribed an operator $A$, acting in $\cal E$; such
quantities are called {\it observables}.
\jt{3.} Only the eigenvalues of the operator $A$ are possible results of
a single measurement of the corresponding observable $\cal A$.
\jt{4.} When the observable $\cal A$ is measured on a system in the state
$\ket{\j}$, the probability ${\cal P}(a_n)$ of obtaining the
non-degenerate\ft{Degenerate generalizations are easy and merely
technical, not of principle.} eigenvalue $a_n$ of the corresponding
operator $A$ is
\eqn\eXXX{ {\cal P}(a_n)~ = ~{\big|\V{u_n|\j}\big|^2\over\V{\j|\j}}~,
            \qquad A\ket{u_n}~ = a_n\ket{u_n}~, }
\ie, $\ket{u_n}$ is the normalized eigenstate of $A$ associated to the
eigenvalue $a_n$
\jt{5.} If the measurement of the physical quantity $\cal A$ on the
system in the state $\ket{\j}$ gives the result $a_n$, the state of the
system upon the measurement is the normalized projection
\eqn\eXXX{ {P_n\ket{\j} \over \sqrt{\V{\j|P_n|\j}}} ~=~ \ket{u_n} }
of $\ket{\j}$ onto the eigenstate associated with $a_n$.
\jt{6.} The time evolution of the state vector $\ket{\j}$  is
governed by the Schr\"odinger equation:
\eqn\eSch{ i\hbar\>{\rd\over\rd t}\ket{\j}~ = ~H\ket{\j}~, }
where $H$ is the Hamiltonian operator of the system.\ping

It is usually {\it implicitly assumed} that the state set $\cal E$
from~{\bf1}, also being the solution set of Eq.~\eSch\ from~{\bf6}, is a
vector space. Equivalently, one regards the Schr\"odinger equation~\eSch\
as being linear---just as it appears to be: Solutions of a linear equation
form a vector space since the sum of any two solutions is again a solution;
this then is tantamount to the `superposition principle'. Note that this
is actually neither included, nor strictly a consequence of the above
axioms. To make this implicit assumption manifest, we remove the seventh
veil:
 \vglue2pt\noindent
\jt{7.} The Hamiltonian $H$ in axiom~{\bf6} is independent of $\ket{\j}$.
 \vglue2pt\noindent
As will hopefully become evident below, the assertion of~{\bf7} should not
be regarded as another axiom, but merely as an
interpretational/applicational {\it choice}; in fact, its negation
($\overline{\bf7}$) is equally viable and perhaps even more interesting
(see below).

Contrasting the linearity of Eq.~\eSch, the projection (collapse) in
axiom~{\bf5} is discontinuous, arguably non-linear; this
discrepancy is then argued to in fact imply an inconsistency of quantum
mechanics and to be at the heart of the of the ``quantum measurement
problem''~\rAlbert\ft{In Ref.~\rAlbert, axiom~{\bf6} is labeled `C' and
axiom~{\bf5} is `E'.}. Whilst this latter observation remains to seem
true, the claimed inconsistency turns out to be a mirage --- owing in part
to the implicit assumption of~{\bf7}.

\newsec{Beguilement Breakdown}\noindent
Standard applications of quantum mechanics machinery are developed upon the
above axioms, and with the (implicit) assertion of~{\bf 7}.
The state vector $\ket{\j}$ is indeed the wave-function that describes the
system under scrutiny, $S$, and so is assumed to carry all possibly
knowable information about it. The differential equation~\eSch\ does
embody the dynamical principle which determines the time-evolution (and
all other characteristics) of the state vector $\ket{\j}$ and so provides
a complete description of the quantum dynamics of $S$.

However, just what {\it is} $H$? Textbooks prescribe how to determine the
{\it operator} $H$, which typically looks something like $H=T+V$, where $T$ is
the kinetic energy (operator) of the system and $V$ the potential energy
(operator). The kinetic operator $T$ determines the evolution of the state
vector $\ket{\j}$ (and so the system $S$) in lieu of any interaction,
whereas the potential operator $V$ describes the effects of all
interactions affecting $S$.

The physical meaning of the assertion of~{\bf7} is that these interactions
affecting $S$ are due to agents {\it external\/} to $S$,
\ie, the Hamiltonian $H$ specifies the environment {\it external\/} to the
system $S$. This makes explicit the versed practitioners' hitherto mostly
implicit understanding that the Universe is being split asunder into:
\item{1.} $\ket{\j}$, which represents the state of the (sub)system under
scrutiny $S$ and carries all relevant information about it, and
\item{2.} $H$, which represents the `environment' in which $S$
evolves and carries all relevant information about this `environment'.
\vglue0pt\noindent
This is clearly evident even in the wording of most textbook paradigms,
such as `particle in a box', where $\ket{\j}$ represents the particle,
$V$---the box; {\it etc}. On the technical side, the differential
equation~\eSch\ now manifestly {\it is} linear and its solution set, $\cal
E$, therefore necessarily {\it is} a vector space --- the superposition
principle is applicable.
\bigskip

This division is very well suited for {\it typical} applications of
quantum mechanics: for modeling of processes in which the scrutinized
(sub)system is usually well (qualitatively and especially quantitatively)
distinguished from the environment. For example, in the classic
beam-splitting experiment, the electron beam is split in two, one with
`spin-up' and the other with `spin-down'. The electron beam, as described
by its ket
$\ket{\j}$, is perfectly clearly distinguished from the magnets---which
produce the magnetic field (environment) that interacted with the beam and
caused the splitting. This magnetic field (and so the magnet producing it)
is of course described by an appropriate term in the Hamiltonian $H$.

\subsec{Measurement milieu}\subseclab\sMeasure\noindent
Now, the act of measurement is itself a form of interaction, patently of
the system under scrutiny, $S$, with the measuring machine, $M$.
Therefore, it ought to be possible to describe such an interaction by
including an appropriate `potential'---the one which contains all the
information about the measuring machine, $M$, including the time when it
is set to measure. 

In the `interaction picture', the Schr\"odinger equation~\eSch\ reduces to
\eqn\eSj{ i\hbar{\rd\over\rd t}\ket{\j}~ = ~V_M\ket{\j}~, }
where $V_M$ is the as yet unspecified interaction potential which
describes the interaction of the system under scrutiny, $S$, with
measurement machine, $M$, and contains all the details about the latter.

However, if the measuring machine is itself of this World, and Nature
really is Quantum, then there ought to exist a set of state vectors,
$\ket{\f}$, which describe the quantum states of the measuring machine.
The dynamics of these $\ket{\f}$'s then ought to be determined by another
Schr\"odinger equation, and in the `interaction picture', we have:
\eqn\eSf{ i\hbar{\rd\over\rd t}\ket{\f}~ = ~\L_M\ket{\f}~, }
where the operator $\L_M$ describes (among other things also) how the
measuring interaction with $\ket{\j}$ (re)acts on the measuring device.

Finally, what can one say {\it in general} of the measurement interaction
operators $V_M$ and $\L_M$? A moment's reflection will satisfy the Reader
that $V_M$ {\it must} depend on the state of the measuring machine,
$\ket{\f}$, and likewise that $\L_M$ {\it must} depend on the state of the
(sub)system under scrutiny, $\ket{\j}$. Therefore, the two equations~\eSj\
and~\eSf\ may be written a bit more explicitly as
\eqna\eSys
 $$\eqalignno{
 i\hbar{\rd\over\rd t}\ket{\j}
       &= V_M(\f)\ket{\j}~, &\eSys{a}\cr
 i\hbar{\rd\over\rd t}\ket{\f}
       &= \L_M(\j)\ket{\f}~. &\eSys{b}\cr
}$$
Owing to the dependence of the `interaction potential' operators
$V_M(\f)$ and $\L_M(\j)$ on the state vectors $\ket{\f}$ and $\ket{\j}$,
respectively, these two differential equations are {\it coupled}. In other
words, the coupled system~\eSys{} is essentially {\it self-referential}.

The bottom line: the system~\eSys{} is non-linear.\ping

A remark is in order. Another description (indeed, {\it the more standard
one}) of a system of two interacting parts is indeed possible where one
does not proceed with a coupled system of equations~\eSys{}. Instead,
describing the {\it combined} scrutinized+measuring system, $S{+}M$, one
introduces a product state vector $\ket{\j,\f}=\ket{\j}\ket{\f}$. This
state vector would again evolve according to a third dynamical
(Schr\"odinger) equation written very much like Eq.~\eSch. In this third
equation, however, the new Hamiltonian would have to be independent of
$\ket{\j,\f}$, and would have to refer to agents {\it external\/} to both
the measuring machine and the scrutinized (sub)system, for the
superposition principle to be applicable (asserting assumption~{\bf7}).
More to the point, however, such a linear description is then in no way
adequate for describing the measurement of the (sub)system $S$ by the
machine $M$. Instead, the combined system $S{+}M$ may now be
(meta-)measured only by agents external to $S{+}M$. This augmentation of
the collection of involved agents then generates the quantum measurement
conundrum: the infinite progression of enlarged measured+measuring systems,
which may be terminated only by eventually including the (unexplained and
undescribed) metaphysical mind of the observer (see also Ref.~\rAuyang).
Instead, the system~\eSys{} appears to be more satisfactorily within the
realm of quantum mechanics.

The above brief analysis presents us with two mutually exclusive
options\ft{You cannot be your cake and eat it too.}:
\item{1.} Quantum mechanics can be arranged to be linear (assert~{\bf7}),
but then cannot describe the measurement process {\it and\/} all the
involved components.
\item{2.} Quantum mechanics can be arranged to describe the measurement
process {\it and\/} all the involved components, but then becomes
non-linear (negate~{\bf7}, \ie, assert~$\overline{\bf7}$).
\bigskip

This exclusion principle seems to be built into the very setting of the
axioms~{\bf1}--{\bf6} together with the (usually {\it implicit\/})
assertion of either the assumption~{\bf7} or its
negative,~$\overline{\bf7}$.
\ping

The second approach (negating assumption~{\bf7}) may seem to lie beyond the
standard applications of quantum mechanics, and for a good reason. Nothing
in the standardly quoted (if quoted at all) axioms~{\bf1}--{\bf6} describes
the measuring device, so one is bound to forge an extension as was done
in subsection~\sMeasure. A guide to this end is provided as much by
heuristic plausibility (see also \SS~D) as by the historical
fact that quantum mechanics already has evolved into (easily non-linear)
quantum field theory. While this evolution happened for totally different
reasons, it may nevertheless be helpful to reconsider old problems from
this newer and perhaps more general vantage point. However, \SS~D
demonstrates that an iterative (adiabatic approximation) approach to
non-linear systems very much like~\eSys{} is in fact standard quantum
mechanics textbook material!

\subsec{A toy model of measuring}\noindent
That the system~\eSys{} is non-linear---and so in general not compatible
with the superposition principle---should be clear form the general theory
of differential equations. This is also easy to see with a simple example:
\eqna\eSYS
 $$\eqalignno{
 i\hbar{\rd\over\rd t}\ket{\j}
       &= a\ket{\f}\ket{\j} ~+~\ldots~, &\eSYS{a}\cr
 i\hbar{\rd\over\rd t}\ket{\f}
       &= b\ket{\j}\ket{\f} ~+~\ldots~, &\eSYS{b}\cr
}$$
where the `interaction potentials' $V_M(\f)$ and $\L_M(\j)$ were expanded,
keeping only the linear terms\ft{Note that the $\ket{\f}$ in the product
$\ket{\f}\ket{\j}$ is to be reinterpreted as an operator acting on
$\ket{\j}$, and {\it vice versa} for Eq.~\eSYS{b}. This reinterpretation
is always possible and becomes trivial in any concrete representation,
where the state vectors are simply wave-functions.}. Including the omitted
higher order terms (represented by the ellipses) is easily seen to only
complicate matters technically, but not in principle.

With the toy model at hand, there are two radically different cases:
1.~when either $a$ or $b$ vanishes, and 2.~when both $a,b$ are nonzero.
(The case when both $a$ and $b$ vanish corresponds to no coupling, that
is, no measurement.)

\bigskip
\leavevmode$\2{b=0}$:~~\ignorespaces
Now Eq.~\eSYS{b} simply states that $\ket{\f}=\f_0$ (in the interaction
picture) does not explicitly depend on time and is determined without ever
asking about $\ket{\j}$. Thereupon, Eq.~\eSYS{a} is indeed linear
in $\ket{\j}$, formally solved by
$\ket{\j}=\exp\big[{-}i{t\over\hbar}a\f_0\big]\ket{\j_0}$.

This result epitomizes the standard practice in quantum mechanics, where
$\ket{\j}$ evolves according to Eq.~\eSYS{a}, $a\f_0$ represents the
unchanging environment (potential) affecting $\ket{\j}$, and $\ket{\j}$
has no effective back-reaction onto the environment. Being
time-independent, $\f_0$ cannot describe a measurement process since the
latter {\it is} discontinuous in time\ft{In another picture where $\f_0$
is discontinuous in time so as to describe a measurement process, the
evolution operator
$\exp\big[{-}i{t\over\hbar}a\f_0\big]$ and so also $\ket{\j}$ will be
discontinuous in time. However, the moment of discontinuity {\it must}
depend on $\ket{\j}$, as the measurement cannot happen when the
(sub)system $S$ is not in the measuring machine $M$---back to the general
(non-linear) case below.}.

This case asserts assumption~{\bf7}, cannot describe the quantum dynamics
of both the scrutinized (sub)system $S$ and the measuring machine $M$, and
is linear.

\bigskip
\leavevmode$\2{a,b\neq0}$:~~\ignorespaces
From the system of two coupled first order differential equations~\eSYS{},
it is always possible to obtain a second order differential equation for
either one of $\ket{\j},\ket{\f}$ uncoupled from the other; this is
sometimes called the integrability condition. Once this differential
equation for, say, $\ket{\j}$ alone is obtained and solved,
$\ket{\f}$ is obtained directly from Eq.~\eSYS{a}, without further
integration:
\eqn\eOther{ \ket{\f} ~=~ i\hbar{\rd\over\rd t}\log\big(\ket{\j}\big)~, }

The integrability condition for $\ket{\j}$ is:
\eqn\eInt{ i\hbar\bigg[\, \ket{\j}{\rd^2\over\rd t^2}\ket{\j} -
                          \Big({\rd\over\rd t}\ket{\j}\Big)^2 -
                          {1\over a}\Big({\rd a\over\rd t}\Big)
                          \ket{\j}{\rd\over\rd t}\ket{\j}\,\bigg]~
          = ~b\ket{\j}^2{\rd\over\rd t}\ket{\j}~, }
and is manifestly non-linear and even non-homogeneous with respect to
rescaling (re-normalizing) the state vector $\ket{\j}$. Therefore, the
solution set of this equation, $\cal E$, is not a linear vector space
and is not compatible with the superposition principle. Since Eq.~\eOther\
determines $\ket{\f}$ given any $\ket{\j}$, solving Eq.~\eInt\ for
$\ket{\j}$ provides a complete solution to the original system~\eSYS{}.

The Reader unsettled by the appearance of non-linear terms like
$\ket{\j}^2$ should note that in {\it any} concrete representation, the
state vector $\ket{\j}$ is replaced by the appropriate wave-function
$\j({\cdots},t)$ and Eq.~\eInt\ becomes a perfectly legitimate, albeit
quite complicated non-linear and non-homogeneous differential equation of
second order. It remains (as always!) to ensure that the resulting
solutions $\j({\cdots},t)$ are normalizable (square-integrable).

This case negates assumption~{\bf7}, does describe the quantum dynamics of
both the scrutinized (sub)system $S$ and the measuring machine $M$ and is
non-linear.

\newsec{Complaints and Conclusions}\noindent
Two complaints to the foregoing discussion and especially the above toy
model come to mind immediately. First, the choice of $b\neq0$ {\it vs}.\
$b=0$ in the toy model (and $\L_M(\j)\neq0$ {\it vs}.\ $\L_M(\j)=0$ in
general) seems to remain up to the Reader. Could it be that the
arbitrariness of the Reader's whim determines whether or not quantum
mechanics is linear?

This complaint is misplaced, for the toy model~\eSYS{} was precisely
that---a toy model, intended merely to demonstrate the intrinsic
nonlinearity. In an actual model, the measurement interaction potentials
$V_M$ and $\L_M$ are completely determined, depending on the details
of the measurement technique and process employed\ft{Should the Reader
wish to champion the claim that measurement really occurs in the
observer's mind, one must then first develop a quantum theory of mind
before the potentials $V_M,\L_M$ can be specified in sufficient detail.
Nevertheless, the present result would seem to persist.}; see \SS~D.

Second, it would appear that the present result abolishes the superposition
principle in quantum mechanics, in face of myriads of experiments which
have in fact brought about the quantum mechanical wave-particle duality.
This again is not so. While appearing as a complaint on principle, this
really is a technical complaint. All experiments---and so also those which
imply the superposition principle---have a finite resolution. It is then
easy to see that the limits on this resolution merely place a limit on the
ratio of (the appropriate matrix elements of) $V_M$ and $\L_M$. So, while
(this version of) quantum mechanics is intrinsically non-linear, it may
well be negligibly so. In this sense, the present result extends rather than
abolishes the hitherto known quantum mechanics.

In addition, note that $b$ in~\eSYS{b} and $\L_M(\j)$ in general~\eSys{b}
are really functions of time. If chosen so that they (well-nigh) vanish
except at the time of measurement when they are non-negligible, the
cherished linearity of quantum mechanics is (well-nigh exactly)
recovered---except during the event and at the `location'\ft{Here,
`location' refers to a subdomain in space for coordinate representation,
in momentum-space for the momentum representation, etc.} of measurement. 
\ping

The erudite Reader will have realized that the simple toy model~\eSYS{} is
but a special case of the `predator-pray' system, and that~\eSys{}
generalizes this considerably. These systems being non-linear, there
definitely exist whole uncharted worlds of chaotic regimes and effects.
Note, however, that these are radically different (and presumably wilder)
than the effects studied in what is called `quantum chaos'.

In any case (cf.\ Ref.~\rAlbert), the non-linearity of the process of the
``wave-function collapse'' or ``state vector projection'' (axiom~{\bf5})
is now seen to be perfectly consistent with the dynamical evolution
law~\eSch\ (axiom~{\bf6}).

\newsec{A Diatomic Drill}\noindent
Non-linearity is rather commonplace in practically all (classical
and quantum) field theory, and so for this audience the present results
may appear unsurprising. However, to the best of understanding of the
present author, it is not widely popularized that such intrinsic
non-linearity naturally extends (descends?) to quantum mechanics. The
purpose of this section is to demonstrate that non-linearity is really not
novel in quantum mechanics either; the outright statement of this
non-linearity however {\it is}.

Consider a once ionized Hydrogen molecule. Following a standard
textbook~\rPark~(\SS~18.1), we first assume that the two protons are
at a fixed distance $R$, and let $r_a$ and $r_b$ denote the distances of
the electron from one and the other proton, respectively. The
Schr\"odinger equation for the state vector of the electron, $\ket\j$, is
then
\eqn\eXXX{ - \Big[{\hbar^2\over2m} \vec{\nabla}^2 +
  {e^2\over4\p\e_0}\Big({1\over r_a}+{1\over r_b}\Big)\Big]\ket{\j}
           ~=~ E_{\rm elec.}\ket{\j}~, }
or
\eqn\eElec{ H_{\rm elec.}\ket{\j}~\define~
            - \Big[{\hbar^2\over2m} \vec{\nabla}^2\j +
            {e^2\over4\p\e_0}\Big({1\over r_a}
           {+}{1\over\sqrt{r_a^2{+}R^2{-}2r_aR\cos\q}}\Big)\Big]\ket{\j}
           ~=~ E_{\rm elec.}\ket{\j}~, }
where $\q$ is the angle between $\vec{R}$ and $\vec{r}_a$.

Of course, the distance $R$ between the two nuclei is not fixed, but is
an observable, to be calculated as an expectation value and using the
state vector $\ket\vf$ that describes the state of the two protons. For
the latter one, there will then exist a Schr\"odinger equation of the
general type
\eqn\eNucl{ H_{\rm nucl.}\ket{\vf}~\define~\Big[
 -{\hbar^2\over2M}\big(\vec{\nabla}^2_1+\vec{\nabla}^2_2\big)
 +{2e^2\over4\p\e_0}{1\over |\vec{r}_1-\vec{r}_2|}
 +V_{\rm elec.}(R)\Big]\ket{\vf} ~=~ E_{\rm nucl.}\ket{\vf}~, }
where $\vec{r}_i$ is the position vector of the $i^{th}$ nucleus,
$\vec{\nabla}^2_i$ the Laplacian with respect to the coordinates of the
$i^{th}$ nucleus.

Finally, note that
\eqn\eCpl{ V_{\rm elec.}(R) ~\define~ \V{\j|H_{\rm elec.}|\j}~,\qquad
           R ~\define~
          \big\langle\vf\big||\vec{r}_1-\vec{r}_2|\big|\vf\big\rangle~, }
which clearly provide a {\it non-linear} coupling between Eqs.~\eElec\
and~\eNucl. Also, one may interpret this set-up as the two nuclei,
represented by $\ket{\vf}$, `observing' or `measuring' (certain
characteristics of) the electron, represented by $\ket{\j}$, and the other
way around.

In the linear approach, as in Ref.~\rPark, \SS~18.2, one introduces a
total state vector, $\ket{\J(\vec{r}_1,\vec{r}_2,\vec{r}_{\rm el.})}$, for
the 3-body system as a whole, and use the total Hamiltonian
\eqn\eHAM{ {\eqalign{H_{\rm tot.}
 &=~ -{\hbar^2\over2M}\big(\vec{\nabla}^2_1+\vec{\nabla}^2_2\big)
     +{2e^2\over4\p\e_0}{1\over |\vec{r}_1-\vec{r}_2|} \cr
 &\mkern60mu-{\hbar^2\over2m}\big(\vec{\nabla}^2_{\rm el}\big)
  -{2e^2\over4\p\e_0}\Big({1\over |\vec{r}_1-\vec{r}_{\rm el}|}
                +{1\over |\vec{r}_2-\vec{r}_{\rm el}|}\Big)~.\cr}}}
which is independent of $\ket\J$. In this approach, the 3-body system is
an indivisible whole and the Schr\"odinger equation with the
Hamiltonian~\eHAM\ is linear in $\ket\J$. Any \hbox{(meta-)measure}ment of
this 3-body system must be done by an external agent (meta-observer), with a
corresponding term added to the Hamiltonian. This meta-observer itself
however remains undescribed by quantum mechanics until either a non-linearly
coupled system of equations akin to~\eSys{} is given, or $\ket{\J}$ is
extended by another factor for this meta-observer. In the latter case, one
needs a meta-meta-observer to collapse this new wavefunction\3

This hopefully convinces the Reader that quantum mechanics can either be
linear or describe the measurement process, but never both simultaneously.

\vskip0pt plus.2\vsize\penalty-250\vskip0pt plus-.2\vsize
\vfill
\footatend\vfill\supereject\immediate\closeout\rfile\writestoppt
\baselineskip=14pt\centerline{{\bf References}}\bigskip{\frenchspacing%
\parindent=20pt\escapechar=` \input refs.tmp\vfill\eject}\nonfrenchspacing

 %
\bye